\documentclass[prb,aps,twocolumn,amsmath,amssymb,floatfix,
superscriptaddress]{revtex4}
\usepackage[dvips]{graphics}
\usepackage{graphicx}
\usepackage{color}
\definecolor{dred}{rgb}{0.75,0,0}

\begin{document}

\title{{Flux driven and geometry controlled spin filtering for arbitrary spins in aperiodic quantum networks}}

\author{Amrita Mukherjee}
\email{amritaphy92@gmail.com}
\author{Arunava Chakrabarti}
\email{arunava_chakrabarti@yahoo.co.in}
\affiliation{Department of Physics, University of Kalyani, Kalyani,
West Bengal-741 235, India}
\author{Rudolf A.\ R\"{o}mer}
\email{r.roemer@warwick.ac.uk}
\affiliation{Department of Physics and Centre for Scientific Computing, 
University of Warwick, Coventry CV4 7AL, U.K.}

\begin{abstract}
We demonstrate that an aperiodic array of certain quantum networks comprising magnetic and non-magnetic atoms can act as perfect spin filters for particles with arbitrary spin state. This can be achieved by introducing minimal quasi-one dimensionality in the basic structural units building up the array, along with an appropriate tuning of the potential of the non-magnetic atoms, the tunnel hopping integral between the non-magnetic atoms and the backbone, and, in some cases, by tuning an external magnetic field. This latter result opens up the interesting possibility of designing a flux controlled {\it spin de-multiplexer} using quantum networks. The proposed networks have close resemblance with a family of recently developed photonic lattices, and the scheme for spin filtering can thus be linked, in principle, to a possibility of suppressing any one of the two states of polarization of a single photon, almost at will. We use transfer matrices and a real space renormalization group scheme to unravel the conditions under which any aperiodic arrangement of such topologically different structures will filter out any given spin projection. Our results are analytically exact, and corroborated by extensive numerical calculations of the spin polarized transmission and the density of states of such systems. 
\end{abstract}

\maketitle

\section{Introduction}
Spintronics is all about implementing the idea of transporting information through the electron's spin, instead of its charge.\cite{Prinz1995,Prinz1998,Wolf2001} Naturally, the need to gain a comprehensive control over the prospect of filtering out one component (projection) of the two spin states of an electron and generating a spin polarized current turns out to be an important issue in developing spintronic devices.\cite{Murakami2003}
%
Experiments, beginning a couple of decades ago, exploited the 
quantum confinement of electrons\cite{Goldhaber-Gordon1998,Cronenwett1998} and the 
tunability of spin filters in GaAs samples was studied in detail.\cite{Rokhinson2004} The development of a quantum spin pump using a GaAs quantum dot (QD),\cite{Watson2003} and spin polarized transport studies in magnetic nanowires\cite{Rodrigues2003} ushered new light into this exciting research arena. One should also mention molecular wires and spin polarized tunneling device,\cite{Andres1996} which were also examined before as potential candidates to achieve spin controlled transport.

The experiments inspired a lot of theoretical investigations which revealed interesting properties related to spin transport and filtering in quantum devices. These systems do not remain far from being realized in real life, thanks to the immense advancement in lithography and nano-technology. To name a few such theoretical studies, spin filtering and complete localization effect in a QD network,\cite{Bercioux2005} or the interplay of Rashba spin orbit interaction (RSO) and an external magnetic field, leading to a spin filtering effect in a QD network,\cite{Aharony2008,Foldi2008} were among the earlier investigations. Spin polarized coherent electronic transport in low dimensional networks of QD's or magnetic nanowires,\cite{Lu2011,Wang2006b,Shokri2005,Mardaani2006,Dey2011} or the study of a silicine nanoribbon\cite{Nunez2016a} and spin filtering in an engineered graphene nano-ribbon\cite{Kang2017} enrich the recent literature, revealing many subtleties in spin polarized quantum transport.

The quantum network devices (QND) modeled as described above have multiple loop structures providing a variety of quantum interference effects which are crucial in designing spin filters. Even in simple forms, QND's, described within a tight binding framework and without consideration of the RSO interaction have been shown to lead to spin filtering effects.\cite{Fu2012} 

\begin{figure}[tb]
\includegraphics[width=\columnwidth]{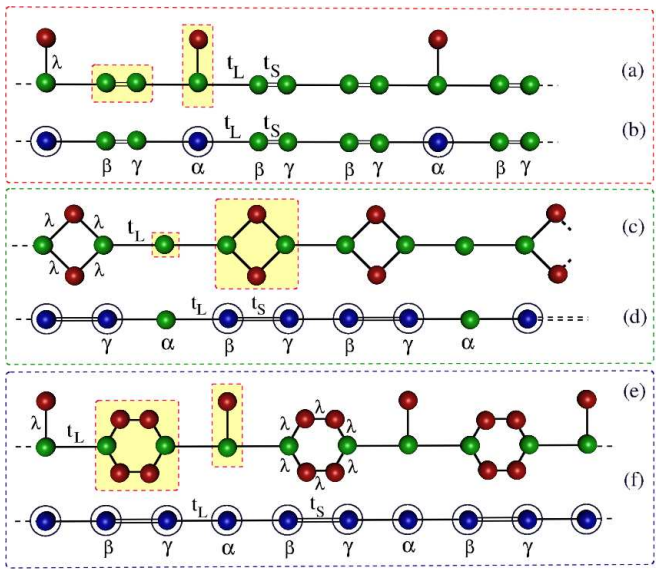}
\caption{(Color online). Three examples of quasiperiodic Fibonacci sequence of quantum network units with different geometries are depicted in (a), (c) and (e). The basic building blocks (highlighted) in each array  consist of magnetic (green) sitting on the backbone, and non-magnetic (red) atoms coupled to them from one side, as shown. Figures (b), (d) and (f) represent the effective linear chains that are obtained by renormalization of the structural units, as explained in the text.} 
\label{fig-lattice1}
\end{figure}

The theoretical work done so far is confined mainly to spintronics for electrons. Only recently an idea of having a spin filter for higher spins by engineering the substrate, composed of a periodic array of magnetic atoms, was proposed and analysed in details.\cite{Pal2016e} To the best of our knowledge, no results exist which explore the possibility of observing spin polarized transport of a projectile with spin $s \ge 1/2$, when the underlying lattice structure (the QND) is no longer periodic.
To put the issue in a much more direct way, one can simply ask if disorder, which leads to localization of all the single particle states,\cite{Anderson1958} rules out the possibility of spin filtering in a QND. In addition to this, another pertinent question is the role of local topology of the atomic clusters of the QND and the tunability of spin polarized transport by an external agent such as a magnetic field. This paper is our first step to resolve such issues.

We find interesting results. In the first few examples, it is observed that, if a projectile with spin $s \ge 1/2$ travels through a QND constructed as an {\it aperiodic} array of atomic clusters with a  short range hopping, it is very much possible to filter out just {\it one} spin channel out of the available number of $(2s+1)$, blocking the others. This can be achieved by forming the QND  as an essentially linear chain of magnetic atoms, with a set of non-magnetic atoms attached from one side. The system thus attains a quasi-one dimensionality, but at a minimal level. The non-magnetic QD's have to have their on-site potentials tuned to special values, for example, by a gate voltage, to initiate the spin filtering effect. The `special' value of this potential can be calculated exactly. In addition, we show that in certain cases, such filtering can be effected only if the hopping integrals along different branches of the QND have a definite correlation between their numerical values. 
In a second set of examples, we show how, with a prefixed set of values of the parameters of the tight binding Hamiltonian, a wide class of QND's can filter out any desired spin state {\it only} by tuning an external magnetic flux threading the plaquettes of the QND. This tempts us to propose a flux controlled {\it spin demultiplexer} in such low dimensional systems.

It turns out that the spin polarized transport in such aperiodic, quasi 1-d, quantum networks is intimately connected to a complete delocalization of the single particle states under certain {\it resonance} conditions, subtle and unusual. This is a non trivial variation of the canonical case of Anderson localization,\cite{Anderson1958} which has recently been pointed out in the literature,\cite{Pal2014,Pal2013,Nandy2016a} and plays a crucial role in this analysis.

Quite interestingly, one can identify some of the geometries we discuss in this paper, and show in Fig.~\ref{fig-lattice1}, with those developed, in recent times, in the field of photonics.\cite{Real2017,Mukherjee2017,Mukherjee2015} Femtosecond laser writing techniques allow one to build experimentally, `lattices' for light, and that too in various geometries. In scalar-paraxial approximation, the propagation of light in such photonic lattices is governed by a Schr\"{o}dinger type equation.\cite{Mukherjee2017} Comprehensive control can now be achieved over the `inter-site' tunneling and the `on-site' potentials. This makes these systems an ideal test bed for the study of problems related to localization and generation of flat, non dispersive bands in photonic systems, much in the spirit of dealing with spinless fermions on a lattice. It is thus tempting to conjecture that the controlled filtering of one spin projection for a `spin-half' projectile may inspire the idea of suppressing any one of the two states of polarization of a single photon. 

The results we obtain in this communication are valid irrespective of the geometrical nature of the array of the QND's. However, here we present results specifically for a quasiperiodic geometry, viz, QND's in a Fibonacci sequence, which allows us to extract analytically exact results. 
In section \ref{sec-model} we chalk out the scheme of the analysis, and in section \ref{sec-filter} the results are presented without and with a magnetic field through the plaquettes. We conclude in section \ref{sec-conclusions}.

\section{The model and the basic equations} 
\label{sec-model}
Let us refer to the set of QND geometries depicted in Fig.~\ref{fig-lattice1}. We first explain the scheme in terms of the simplest looking system, which is Fig.~\ref{fig-lattice1}(a), henceforth referred to as the `dot-stub' chain. Its an electronic counterpart of a similar dot-stub photonic lattice, that was fabricated by laser inscription and investigated by Real et al.\cite{Real2017} who demonstrated there that the trapped photonic modes in phase coherent superpositions lead to all optical logic gate operations.
In the spin filtering problem discussed here, a sequence of magnetic atoms is arranged in a quasi-periodic Fibonacci sequence. We have two kinds of {\it bonds}, namely, $L$ (for `long', say) and $S$ (for `short'), marked by a `double' bond. The chain grows following the well known Fibonacci inflation rule $L \rightarrow LS$ and $S \rightarrow L$, and begins with an $L$ bond. We work within a tight-binding formalism, and the Hamiltonian is given by,
\begin{eqnarray}
\mathbf{H} = \sum_n \mathbf{c}^{\dag}_n \left(\mathbf{\epsilon}_n -\mathbf{h}_n \cdot {\mathbf{s}}_n \right) \mathbf{c}_n +
\sum_{\langle n,m\rangle} \left(\mathbf{c}^{\dag}_n \mathbf{t}_{n,m} \mathbf{c}_m + 
h.c.\right)
\label{ham}
\end{eqnarray}
with $\langle {n,m}\rangle$ denoting nearest neighbors. 
Each of the operators $\mathbf{c}^{\dag}_n$ and $\mathbf{c}_n$,
is a single column or row with the number of entries depending on the
spin component. For example, for a spin-half particle, the creation (annihilation) operator $\mathbf{c}^{\dag}_n$ 
($\mathbf{c}_n$), the on-site energy matrix $\mathbf{\epsilon}_n$, and the nearest neighbor hopping matrix $\mathbf{t}_{n,m}$ are 
\begin{eqnarray}
\mathbf{c}^{\dag}_n = 
\left( \begin{array}{cccc}
c^{\dag}_{n,\uparrow} & c^{\dag}_{n,\downarrow}  
\end{array}
\right) \nonumber, \quad\hfill
\mathbf{c}_n = 
\left( \begin{array}{cccc}
c_{n,\uparrow}  \\ 
c_{n,\downarrow}  
\end{array}
\right), \nonumber \\
\mathbf{\epsilon}_n =
\left( \begin{array}{cccc}
\epsilon_{n,\uparrow} 
& 0 \\                  
0 & \epsilon_{n,\downarrow}
\end{array}
\right), \quad\hfill
\mathbf{t}_{n,m}=
\left( \begin{array}{cccc}
t_{n,m} 
& 0 \\                  
0 & t_{n,m}
\end{array}
\right) .
\label{matrices}
\end{eqnarray}
The term $\mathbf{h}_n \cdot \mathbf{s}^{(s)}_n = h_{n,x} s^{(s)}_{n,x}
+ h_{n,y} s^{(s)}_{n,y} + h_{n,z} s^{(s)}_{n,z}$ in \eqref{ham} describes the interaction of the spin ($s$) of the incoming projectile with the localized on-site magnetic moment $\mathbf{h}_n$ at site \textit{n}. For spin-half, the explicit form of $\mathbf{h}_n \cdot \mathbf{s}^{(s)}_n$ in terms of a matrix representation is given by,\cite{Pal2016e}
\begin{eqnarray}
\mathbf{h}_n \cdot \mathbf{s}^{1/2}_n =  
\left( \begin{array}{cccc}
h_n \cos{\theta_n} &  h_n \sin{\theta_n} e^{-i\phi_n}\\ 
h_n \sin{\theta_n} e^{i\phi_n} & -h_n \cos{\theta_n} 
\end{array}
\right) ,
\label{interact}
\end{eqnarray}
where $h_n$, $\theta_n$ and $\phi_n$ represent the radial component and the polar and azimuthal angles, respectively.

The Fibonacci arrangement of the bonds requires separate nomenclature for the on-site potentials. We assign the names as follows. The site flanked by an $LL$ pair is named $\alpha$, while the sites sitting in between an $LS$ and an $SL$ pair of bond are named $\beta$ and $\gamma$ respectively. There is a single level non-magnetic QD (cp. Fig.~\ref{fig-lattice1}(a)) attached from one side to every $\alpha$ vertex. The tunnel hopping integral between the dot and the backbone is termed $\lambda$. 
In the analysis that follows, we set the on-site potential at each site on the backbone as $\epsilon_{\alpha,\sigma}=\epsilon_{\beta,\sigma}=\epsilon_{\gamma,\sigma}=\epsilon$, for every spin projection $\sigma$. Its understandable that $\sigma=1/2$ ($\uparrow$), or $-1/2$ ($\downarrow$) in the spin-half case, while, $\sigma=1$, $0$ and $-1$ for a particle with total spin $s=1$, and so on. The side-coupled QDs in Figs.\ \ref{fig-lattice1}(a+c+e) are non-magnetic in nature, and are assigned a potential $\epsilon_N$, that can be tuned by a gate voltage.
The strength of the magnetic moment (equivalently, the `local field') $h_n$ can, in principle, assume three different values, viz, $h_\alpha$, $h_\beta$ and $h_\gamma$ for the $\alpha$, 
$\beta$ and $\gamma$ sites respectively, depending on the chemical species of the atoms employed. For simplicity we choose $h_\alpha=h_\beta=h_\gamma=h$ in what follows here. 

We calculate transmission properties for different spin channels using the standard transfer matrix method, assuming that two semi-infinite, perfectly periodic and non-magnetic leads connect the system at its left and the tight ends. The leads are described by a tight binding Hamiltonian, and have on-site potential $\epsilon_\mathrm{lead}$, and nearest neighbor hopping integral $t_\mathrm{lead}$. The method is discussed in further detail elsewhere.\cite{Pal2016e}

\section{Spin filtering without external magnetic field}
\label{sec-filter}

\subsection{The spin-half case and the dot-stub geometry}

To explain the basic scheme, we choose the `dot-stub' geometry in Fig.~\ref{fig-lattice1}(a), and the spin-half case at the beginning. 
We choose $\theta_n=\phi_n=0$ to first unravel the spin filtering properties in a completely analytical way. We begin by decimating the amplitude of the wave function at every non-magnetic site, in terms of the amplitude at the $\alpha$ site at its base. Once this is accomplished, the amplitude of the wave function at site $n$, lying entirely on the backbone, satisfies the Schr\"{o}dinger equation $H\Psi = E \Psi$, with $\Psi=\sum_{n,\sigma} \psi_{n,\sigma} |n,\sigma \rangle$ , and $\sigma=\pm 1/2$, written in an equivalent `difference equation' form as\cite{Pal2016e}

\begin{multline}
\left\{E-\left(\epsilon - 2\sigma h + \frac{\lambda^2}{E-\epsilon_N}\right )\right\}\psi_{n,\sigma}
\nonumber
\end{multline}
\begin{align}
& = & t_L \psi_{n-1,\sigma} + t_L \psi_{n+1,\sigma} , \nonumber \\
\left[ E - (\epsilon - 2\sigma h) \right] \psi_{n,\sigma} 
& = & t_L \psi_{n-1,\sigma} + 
t_S \psi_{n+1,\sigma} , \nonumber \\
\left [E - (\epsilon - 2\sigma h) \right] \psi_{n,\sigma} 
& = & t_S \psi_{n-1,\sigma} + t_L \psi_{n+1,\sigma} \nonumber \\
\label{diffeqn}
\end{align}

for $\alpha$, $\beta$ and $\gamma$ sites, respectively, and the on-site energy at the non-magnetic sites is denoted by $\epsilon_N$.
%
%
It is interesting to note that, this seemingly trivial one dimensional backbone has the flavor of an {\it extra} or {\it synthetic} dimension hidden in it, which unfolds only to the incoming projectile depending on its spin state $s$. The array of magnetic atoms appears as a $(2s+1)$-strand {\it ladder network} to a projectile with spin $s$.\cite{Pal2016e}
Incidentally, similar multi-strand ladder networks (MLN) in tight-binding formalism have previously been explored as prototypes of DNA molecules, with the inter-arm `cross hoppings' along the diagonals\cite{Paez2012b} simulated here by the  terms $h_n \sin\theta_n e^{\pm i\phi_n}$ in Eq.~\eqref{interact}, in respect of their device aspects or charge transportation.\cite{WelCR09,Cuniberti2007f} Some other studies involving similar MLN's include the issue of delocalization of single particle states in properly engineered disordered or aperiodic quantum networks.\cite{Sil2008a,Rodriguez2012}

\subsection{Engineering a spin filter}

Eq.~\eqref{diffeqn} is actually a set of six equations, grouped in two subsets. Each subset, consisting of {\it three} equations, represents two decoupled, independent Fibonacci chains. In each subset, the first equation is written for an $\alpha$ site, while the two subsequent equations are written for the sites of type $\beta$ and $\gamma$ respectively. The $\alpha$-site potentials for the $\uparrow$ and $\downarrow$ spin projections for the two decoupled Fibonacci chains are given, respectively, by, 

\begin{eqnarray}
\tilde\epsilon_{\alpha,\sigma} & = & \epsilon -2\sigma h + \frac{\lambda^2}{E-\epsilon_N}  ,
\label{alphasite}
\end{eqnarray}
while for the $\beta$ and $\gamma$ sites these are $\tilde\epsilon_{\beta,\sigma}=\tilde\epsilon_{\gamma,\sigma}=\epsilon -2\sigma h$.
A pertinent issue to discuss here, is the role of the `local' magnetic field $h$ offered by the magnetic atoms on the backbone. A large value of $h$ will naturally split the bands for the $\uparrow$ and the $\downarrow$ spins.\cite{Pal2016e} Therefore, even when $\lambda=0$, that is, when we have a purely one dimensional Fibonacci lattice, the $\uparrow$ and $\downarrow$ spins will have their spectra separated on the energy scale. Each such spectrum will have the usual three subband structure\cite{Kohmoto1987}. The transport for the two spin projections will be there, over these two energy regimes, exhibiting the usual multifractal character\cite{Kohmoto1987}, thinning out as the system attains its thermodynamic limit. Spins will still get filtered out, but in a scanty, fractal way. Most importantly, due the Cantor set character of the energy spectrum, it is impossible to locate an energy eigenvalue exactly for an infinite system. 

In this paper, we engineer absolutely continuous bands in such a quasiperiodic arrangement of structural units, and obtain continuous and {\it completely unattenuated} spin transport, filtered for $\uparrow$ and $\downarrow$ spins at appropriately chosen domains over the full spectral zone.
\begin{figure}[bt]
\includegraphics[width=\columnwidth]{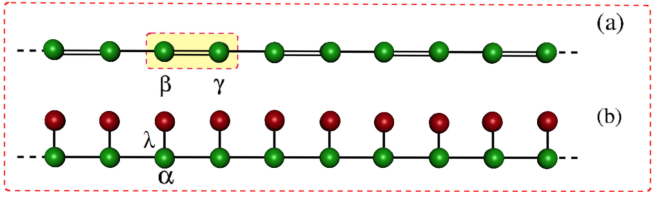}
\caption{(Color online).(a) The periodic, infinitely long $\beta\gamma$ dimer (shown in dotted box) lattice, and (b) The infinite periodic array of the `stubbed' $\alpha$ sites. The on site potentials are, $\epsilon_\alpha=\epsilon_\beta=\epsilon_\gamma=\epsilon_N=\epsilon\mp h$ for the $\uparrow$ and $\downarrow$ spins. The vertical `tunnel' hopping in (b) is chosen as $\lambda=\sqrt{t_S^2-t_L^2}$ for the matrices to commute. The densities of states for these two lattices merge as the commutation of the transfer matrices is enforced. The colors are as in Fig.\ \ref{fig-lattice1}.}
\label{fig-decoupledchains}
\end{figure}
%
Let us look at Fig.~\ref{fig-lattice1}(b). On this effectively one dimensional Fibonacci chain, we find two distinct `building blocks', viz, an isolated $\alpha$ site (with renormalized potential), and a `dimer' $\beta\gamma$, arranged following a Fibonacci pattern. This of course, is a generic feature of the Fibonacci lattice grown following the rule stated earlier, and thus remains valid for all the quasi one dimensional quantum networks discussed in this paper. 

Corresponding to two such building blocks, one can construct $2 \times 2$ unimodular `transfer matrices' $\mathcal{M}_{\alpha,\sigma}$ and 
$\mathcal{M}_{\gamma\beta,\sigma} \equiv \mathcal{M}_{\gamma,\sigma} \mathcal{M}_{\beta,\sigma}$, that are given by, 
\begin{eqnarray}
\mathcal{M}_{\alpha,\sigma} & = & 
\left( \begin{array}{cccc}
(E-\tilde\epsilon_{\alpha,\sigma})/t_L & -1 \\ 
1 & 0 
\end{array}
\right), \nonumber \\
\mathcal{M}_{\gamma\beta,\sigma} & = &
\left( \begin{array}{cccc}
\frac{(E-\tilde\epsilon_{\gamma,\sigma})(E-\tilde\epsilon_{\beta,\sigma})}{t_Lt_S}-\frac{t_S}{t_L} 
& -\frac{E-\tilde\epsilon_{\gamma,\sigma}}{t_S} \\                  
\frac{E-\tilde\epsilon_{\beta,\sigma}}{t_S} & -\frac{t_L}{t_S}
\end{array}
\right) .
\label{matrices}
\end{eqnarray}
The on-site potentials, viz, $\tilde\epsilon_\alpha$, 
$\tilde\epsilon_\beta$ or $\tilde\epsilon_\gamma$ assume their 
appropriate values depending on the spin projection, as stated earlier.

For each spin state $\sigma$, the pair of the amplitudes of the wave function at any $n+1$-th and $n$-th sites on the linear backbone is related to any arbitrary pair of sites, marked as $1$ and $0$, for example, through a simple product of $2 \times 2$ transfer matrices
\begin{equation}
\left ( \begin{array}{c}
\psi_{n+1,\sigma} \\
\psi_{n,\sigma}  
\end{array} \right )
= \mathcal{M}_{n,\sigma} \cdot \mathcal{M}_{n-1,\sigma} \cdot \ldots \cdot \mathcal{M}_{2,\sigma} \cdot \mathcal{M}_{1,\sigma} 
\left ( \begin{array}{c}
\psi_{1,\sigma} \\
\psi_{0,\sigma}  
\end{array} \right )
\end{equation}
Let us now work out how to transmit the $\uparrow$ spin for example. We choose the first subset from Eq.~\eqref{diffeqn} corresponding to the $\uparrow$ spin projection, and compute the commutator $\left[\mbox{\boldmath $M_{\alpha,\uparrow}$},\mbox{\boldmath $M_{\gamma\beta,\uparrow}$} \right] $.
The commutator reads,
\begin{widetext}
\begin{equation}
\left[\mathcal{M}_{\alpha,\uparrow},\mathcal{M}_{\gamma\beta,\uparrow} \right] = 
-\frac{(E-\epsilon_N)(t_S^2-t_L^2)-\lambda^2(E-\epsilon+h)}{(E-\epsilon_N)t_Lt_S}
\left( \begin{array}{cccc}
0 & 1  \\ 
1 & 0 
\end{array}
\right)
\label{com}
\end{equation}
\end{widetext}
It is easily verified that, if we set $\lambda=\sqrt{t_S^2 - t_L^2}$, and $\epsilon_N = \epsilon-h$, then $[\mathcal{M}_{\alpha,\uparrow},\mathcal{M}_{\gamma\beta,\uparrow}]=0$, {\it independent of energy}. Therefore, in the chosen subset of Eq.~\eqref{diffeqn} that corresponds to the $\uparrow$ spin case, the specific order of arrangement of the pair of sites $\beta\gamma$, and the isolated (stubbed) site $\alpha$ becomes unimportant. 
\begin{figure}[ht]
(a)\includegraphics[width=0.95\columnwidth]{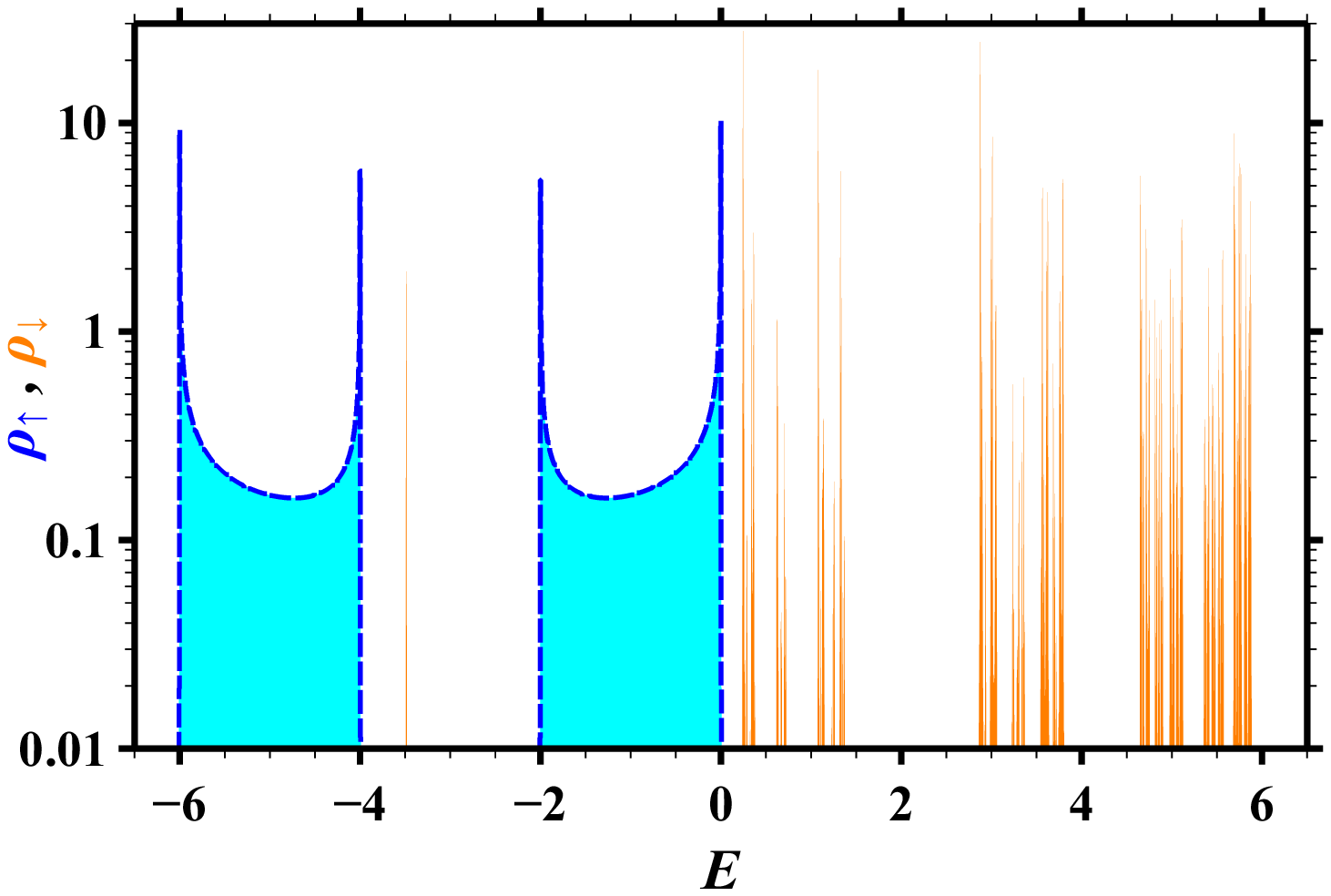}
(b)\includegraphics[width=0.95\columnwidth]{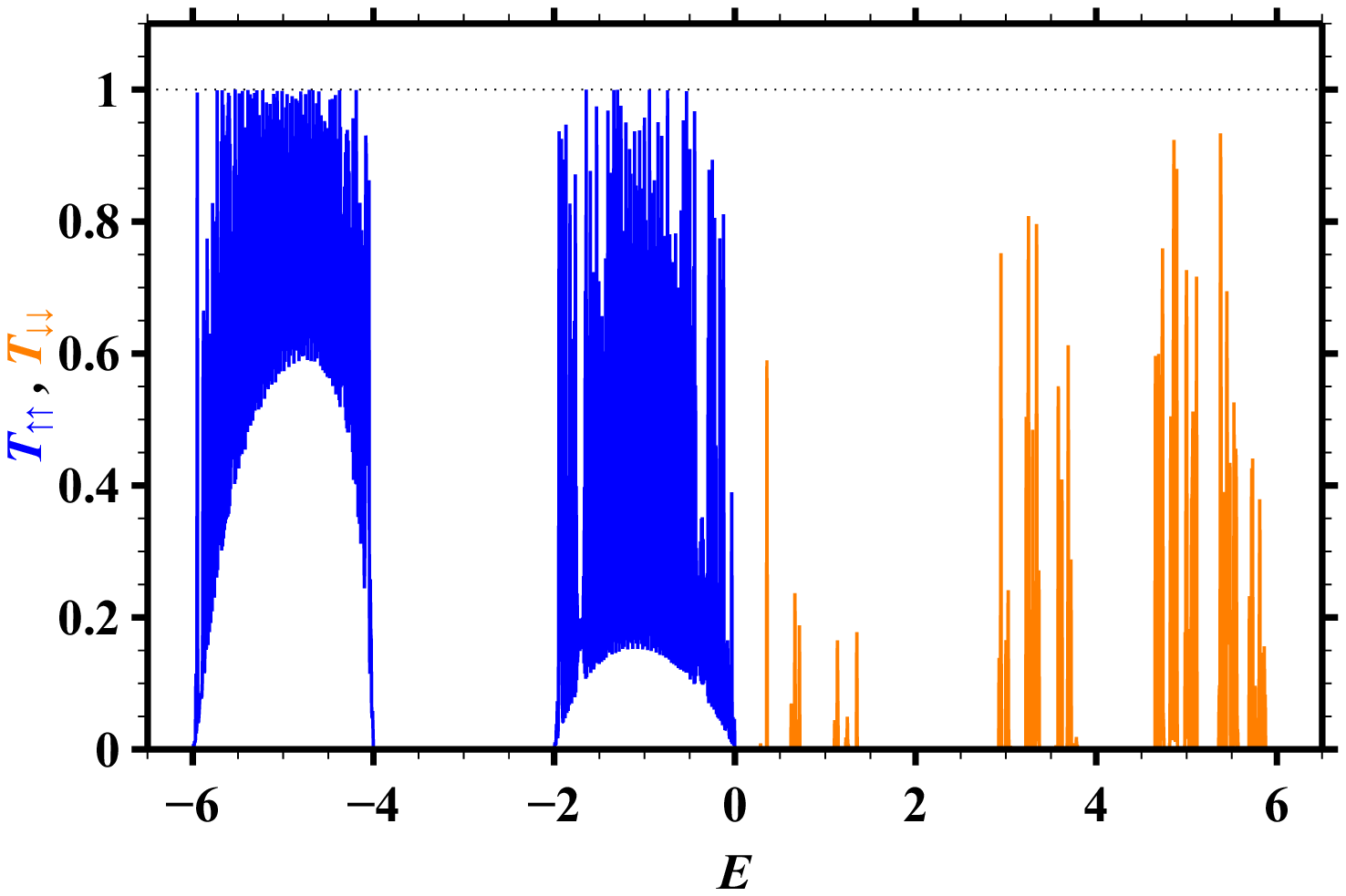}
\caption{(Color online).
(a) DOS for spin-$1/2$ particles in the stub geometry shown in Fig.\ \ref{fig-lattice1} (a+b) for $\epsilon_i=0$, $h=3$, $t_L=1$ and $t_S=2$. The dashed line (with dark/blue shading) indicates the spin-up projection while the solid line (with lighter/orange shading) corresponds to the spin-down case.
(b) The transmission coefficient for the same system and parameters as in (a). The dark solid (blue) line indicates spin-up while the lighter (orange) line is for spin-down. The lead parameters of the non-magnetic leads are $\epsilon_\mathrm{lead}=0$ and $t_\mathrm{lead}=3 t_L$.}
\label{fig-stub}
\end{figure}
Thus, we can, under this `resonance condition', think of the 
Fibonacci array for the $\uparrow$ electrons as being composed of two {\it infinitely long periodic lattices}, one made up of the $\alpha$-sites stubbed with the dots only, and the other, of the pairs $\beta\gamma$, as shown in Fig.~\ref{fig-decoupledchains}. As a result, one expects a complete transparency in the transport of $\uparrow$ electrons over the range of energy $E$ which spans the absolutely continuous spectra offered by these two periodic lattices.

An important point needs to be emphasized here. Each of the two linear periodic lattices in Fig.~\ref{fig-decoupledchains} displays two absolutely continuous subbands in their respective densities of states, which occupy different intervals of energy $E$. As the transfer matrices commute under the special correlations between the potentials and the hopping integrals, as stated above, these two different densities of states {\it have to merge}, and have to become indistinguishable from the DOS of the Fibonacci array of stubbed $\alpha$ sites and the $\beta\gamma$ dimers. Otherwise, the energy interval over which the $\uparrow$ spins will be filtered out is going to be ill defined, and the scheme of spin filtering should not work.

We have checked this analytically by calculating the local densities of states (DOS) $\rho_{\beta,\uparrow}$ at the $\beta$ or $\rho_{\gamma,\uparrow}$ at the $\gamma$ site of the first chain (the periodic $\beta\gamma$ array, with $\rho_{\beta,\uparrow}=\rho_{\gamma\uparrow}$), and $\rho_{\alpha,\uparrow}$ at the $\alpha$ site of the remaining chain. This gives us an estimate of the band positions and the widths in the two cases. We have set, for simplicity of the expressions, $\epsilon_{N,\uparrow}=\epsilon_{\alpha,\uparrow}=\epsilon_{\beta,\uparrow}=\epsilon_{\gamma,\uparrow}=\epsilon-h$ beforehand. The DOS's are given by,
\begin{eqnarray}
\rho_{\beta,\uparrow} & = & \frac{1}{\pi} \frac{E-\epsilon+h}{\sqrt{4t_L^2t_S^2 - [(E-\epsilon+h)^2 - (t_L^2+t_S^2)]^2}} \nonumber \\
\rho_{\alpha,\uparrow} & = & \frac{1}{\pi} \frac{E-\epsilon+h}{\sqrt{4t_L^2 (E-\epsilon+h)^2 - [(E-\epsilon+h)^2-\lambda^2]^2}} \nonumber \\
\label{density}
\end{eqnarray}
It is simple to verify from Eq.~\eqref{density} that, $\rho_{\beta,\uparrow}=\rho_{\alpha,\uparrow}$ as soon as we enforce 
\begin{equation}
\lambda=\sqrt{t_S^2-t_L^2}.
\label{eq-comm}
\end{equation}
The results are similar when we choose to transport the $\downarrow$ spins. The selection of the potentials in this case now will be $\epsilon_{i,\downarrow}=\epsilon+h$, with $i \equiv \alpha$, $\beta$, $\gamma$ and the non-magnetic dot. The choice of the tunnel hopping $\lambda$ remains the same.

It should be appreciated here that the creation of absolutely continuous bands in the DOS spectrum and consequential unattenuated transport is a result of the commutation of the transfer matrices corresponding to two independent constituents (like the highlighted units shown in Fig.~\ref{fig-lattice1}). This happens for any arrangement, including a completely disordered one, of the building blocks shown, and thus presents a non- trivial variation of Anderson localization.\cite{Pal2013} The `order' of arrangement of the units doesn't really matter. This implies that, an infinite variety of geometrical arrangements, periodic, quasiperiodic or random, involving the same networks exhibits complete delocalization of the eigenstates under the same conditions and in a way, group together to exhibit a subtle {\it universality class}. In terms of photonics,  engineering a {\it polarization filter} for photons may be given a consideration, thinking in this line. 
The above discussion remains valid for the two remaining geometries shown in Fig.~\ref{fig-lattice1} as well, for which the difference equations and the commutators are presented in the Appendix. 

Back to the filtering of the  spin states, we see that the second subset in Eq.~\eqref{diffeqn} still represents a quasiperiodic Fibonacci chain for the $\downarrow$ spin electrons, with its own, typically multifractal DOS \cite{Kohmoto1987}. The complete spectrum of the system shown in Fig.~\ref{fig-lattice1}(a) is obtained from a convolution of the DOS arising out of the two subsets of Eq.~\eqref{diffeqn} corresponding to the $\uparrow$ and $\downarrow$ spins. An appropriate choice of the strength of the magnetic moment $h$ can separate out the spectra arising out of the two subsets\cite{Pal2016e} on energy axis, thereby removing the possibility of any overlap between the absolutely continuous sub-bands from the $\uparrow$ spin equations, and the fractal spectrum contributed by the second subset, that is for the $\downarrow$ spins. The $\uparrow$ spin subbands, absolutely continuous in character, should be completely transparent over the range of energy for which $\rho_{\alpha(\beta),\uparrow}$ is non-zero. The $\downarrow$ spins give rise to a multifractal, Cantor set energy spectrum.

The $\downarrow$ spins get transmitted in that part of the energy range, where $\rho_{\alpha(\beta),\downarrow}$ is non-zero. The transmission spectrum is scanty, and one should expect a usual scaling behavior, typical of the Fibonacci lattice, that drops in magnitude as the system grows to its thermodynamic limit. The aperiodic dot-stub array, (and the other ones in Fig.~\ref{fig-lattice1}) can thus act as a perfect spin filter for $\uparrow$ spins. Choosing the self energy of the QD at the stub, as $\epsilon_N=\epsilon+h$, we can have a prefect spin filter for the $\downarrow$ spins, using identical arguments already outlined.

\begin{figure}[tb]
(a)\includegraphics[width=0.95\columnwidth]{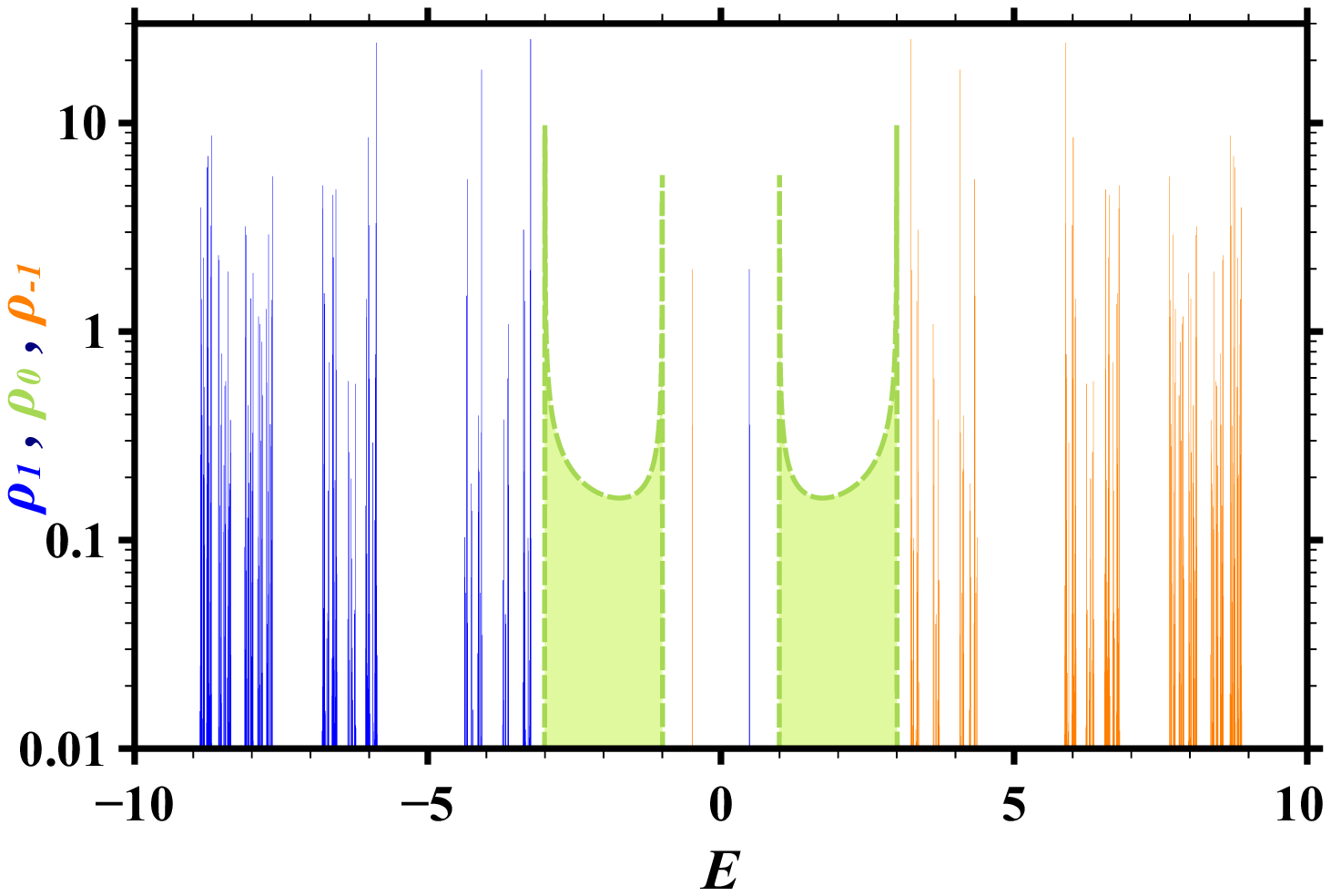}
(b)\includegraphics[width=0.95\columnwidth]{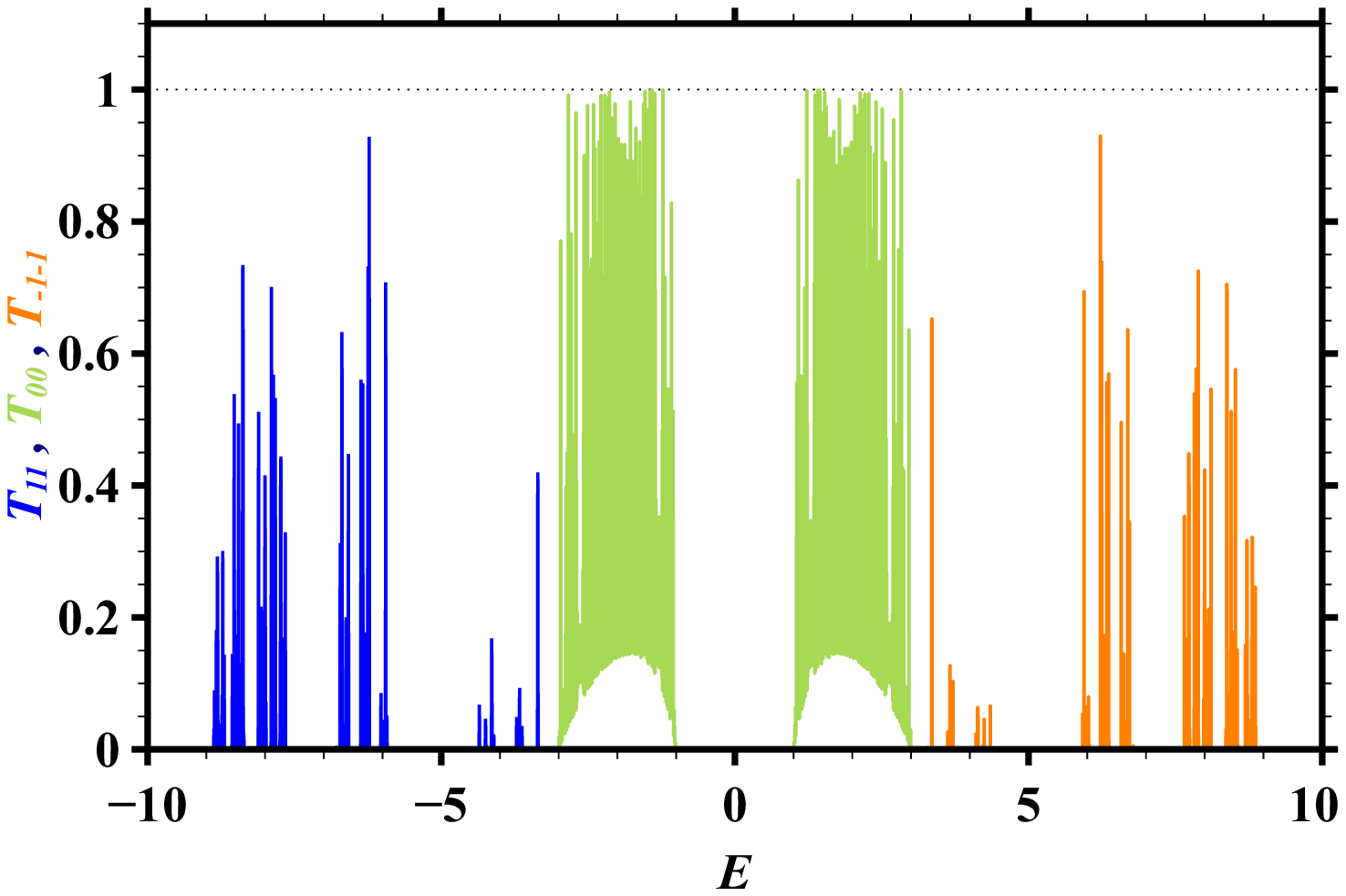}
\caption{(Color online) 
(a) DOS for spin-$1$ particles in the stub geometry shown in Fig.\ \ref{fig-lattice1} (a+b) for $\epsilon_i=0$, $h=3$, $t_L=1$ and $t_S=2$. (b) Transmission coefficient for the stub geometry and parameters as in (a) with lead parameters of the non-magnetic leads given by $\epsilon_\mathrm{lead}=0$ and $t_\mathrm{lead}=5 t_L$. }
\label{fig-spinone}
\end{figure}
In Fig.~\ref{fig-stub} we show the DOS profile and the corresponding transport characteristics of the dot-stub Fibonacci lattice. The DOS has been calculated by evaluating the matrix elements of the local Green's function $\textbf G=(E \bf{1} -\bf H)^{-1}$ for a $377$ bond long lattice. The commutation conditions \eqref{eq-comm} are imposed. 
%
The $\uparrow$ spins exhibit a continuous patch of hight transmission values in the energy regime where the $\uparrow$ spin subbands are absolutely continuous. On the contrary, the $\downarrow$ spin shows scanty, fractal like transmission coefficients in its own `allowed' spectral zones. 

\subsection{Scheme for general spin $s$}

The formalism works perfectly well for any spin $s$. The `virtual' ladder we talked about before, now has $(2s+1)$ strands. If we look at the prospect of spin polarized transport for projectiles with total spin $s$, with $\theta_n=\phi_n=0$ as before, we have a set of $(2s+1)$ decoupled equations. Each such set represents an {\it independent} Fibonacci chain, and is a triplet of equations, corresponding to the sites $\alpha$, $\beta$ and $\gamma$ as its constituents. 
Let us take a specific example. When the spin of the projectile is $s=1$, the spin projections are given by $\sigma=1$, $0$ and $-1$. The on site potential at an $\alpha$ site in the effectively linear Fibonacci chain are, $\tilde\epsilon_{\alpha,\pm 1}=\epsilon \mp h+\lambda^2/(E-\epsilon_N)$, for $\sigma=\pm 1$, and represents the effective potential at a site of type $\alpha$. The $\beta$ and the $\gamma$ sites are crowned with the on-site potential values $\tilde\epsilon_{\beta,\pm 1}=\epsilon_{\gamma,\pm 1} = \epsilon \mp h$.
For the spin projection $\sigma=0$, $\tilde\epsilon_{\alpha,0}=\epsilon + \lambda^2/(E-\epsilon_N)$ for an $\alpha$ site, while $\tilde\epsilon_{\beta,0}=\tilde\epsilon_{\gamma,0} = \epsilon$. The nearest neighbor hopping integrals along the backbone remain as $t_L$ or $t_S$ depending on the bonds. 

Suppose we wish to filter out the spin state $\sigma=0$. For this, we simply need to set $\epsilon_N=\epsilon$.
The relevant transfer matrices for $\sigma=0$ for the dot-stub case in Fig.~\ref{fig-lattice1}(a) now assume the forms, 
\begin{eqnarray}
\mathcal{M}_{\alpha,0} = 
\left( \begin{array}{cccc}
\frac{E-(\epsilon+\frac{\lambda^2}{E-\epsilon})}{t_L} & -1 \\ 
1 & 0 
\end{array}
\right) \nonumber \\
\mathcal{M}_{\gamma\beta,0} =
\left( \begin{array}{cccc}
\frac{(E-\epsilon)^2}{t_Lt_S}-\frac{t_S}{t_L} 
& -\frac{E-\epsilon+h}{t_S} \\                  
\frac{E-\epsilon}{t_S} & -\frac{t_L}{t_S}
\end{array}
\right) 
\label{spinzeromatrices}
\end{eqnarray}
The commutator $[\mathcal{M}_{\alpha,0},\mathcal{M}_{\gamma\beta,0}]=0$ as soon as we set $\lambda$ as in Eq.\ \eqref{eq-comm}. 
This implies that we are going to get absolutely continuous subbands, just as before, corresponding to the the spin state $\sigma=0$. This  particular spin channel will then be completely transparent, while for the two other spin projections, viz, $\sigma=\pm 1$ we shall eventually will get `poor conductance' for a large system. Thus the dot-stub array in this case can be made to act as a spin filter for the $\sigma=0$ state. The selection of $\epsilon_N=\epsilon \mp h$, on the other hand, allows the $\sigma=\pm 1$ states (only one at a time though) to tunnel through, blocking the others.
In Fig.~\ref{fig-spinone} we present the results for $s=1$. With the parameter choices as above, we filter out the spin channel $\sigma=0$ as transmitting, while the other projections $\sigma=\pm 1$, exhibit fractal character in their DOS.
\begin{figure}[tb]
\includegraphics[width=0.95\columnwidth]{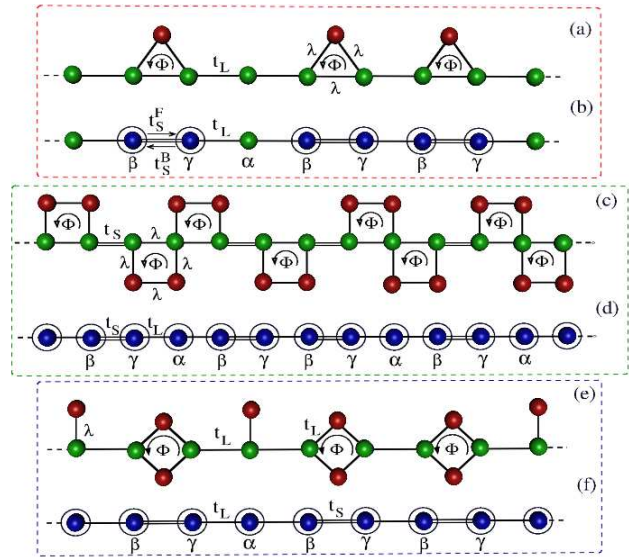}
\caption{(color online). Geometries where magnetic flux plays a pivotal role in spin filtering. (a) A Fibonacci array of triangles and dots, and (b) its renormalized version. (c) A Fibonacci array of diamond shaped plaquettes and stubs, and (d) its effective 
renormalized one dimensional version. (e) and (f) depict the diamond-stub system and the renormalized chain respectively. The colors are chosen as in Fig.\ \ref{fig-lattice1}.}
\label{fig-fluxlattice}
\end{figure}

We end this section bringing an interesting variation of the proposed models to the notice of the reader. The arguments put forward so far for spin-half or spin-one, or, for any spin $s$ will hold perfectly well for a much more general situation.  If the three sites $\alpha$, $\beta$ and $\gamma$ represent three chemically different species with the combinations of the on-site potentials and magnetic moments $(\epsilon_{i,\sigma},h_i)$, with $i \equiv \alpha$, $\beta$ or $\gamma$, even then we can make any desired spin channel transmit, blocking the others. 
For example, considering the spin-half situation, and a target of filtering out the $\uparrow$ spin again, we need to enforce a correlation $\epsilon_N=\epsilon_{i,\sigma} - h_i =$ a constant. One can now afford to take the individual values of $\epsilon_i$ and $h_i$ even from a set of random numbers, but always maintaining the above correlation in their numerical values. The tunnel hopping integral $\lambda$ still should be chosen as $\sqrt{t_S^2-t_L^2}$. The matrices will commute, and we shall have the liberty to engineer a spin filter even now. Same arguments remain valid for any spin state $s$, and for any desired spin projection $\sigma$. The scheme thus goes well beyond a quasiperiodic Fibonacci ordering and encompasses a larger canvas of {\it disordered} systems as well.
\begin{figure}[tb]
(a)\includegraphics[width=0.95\columnwidth]{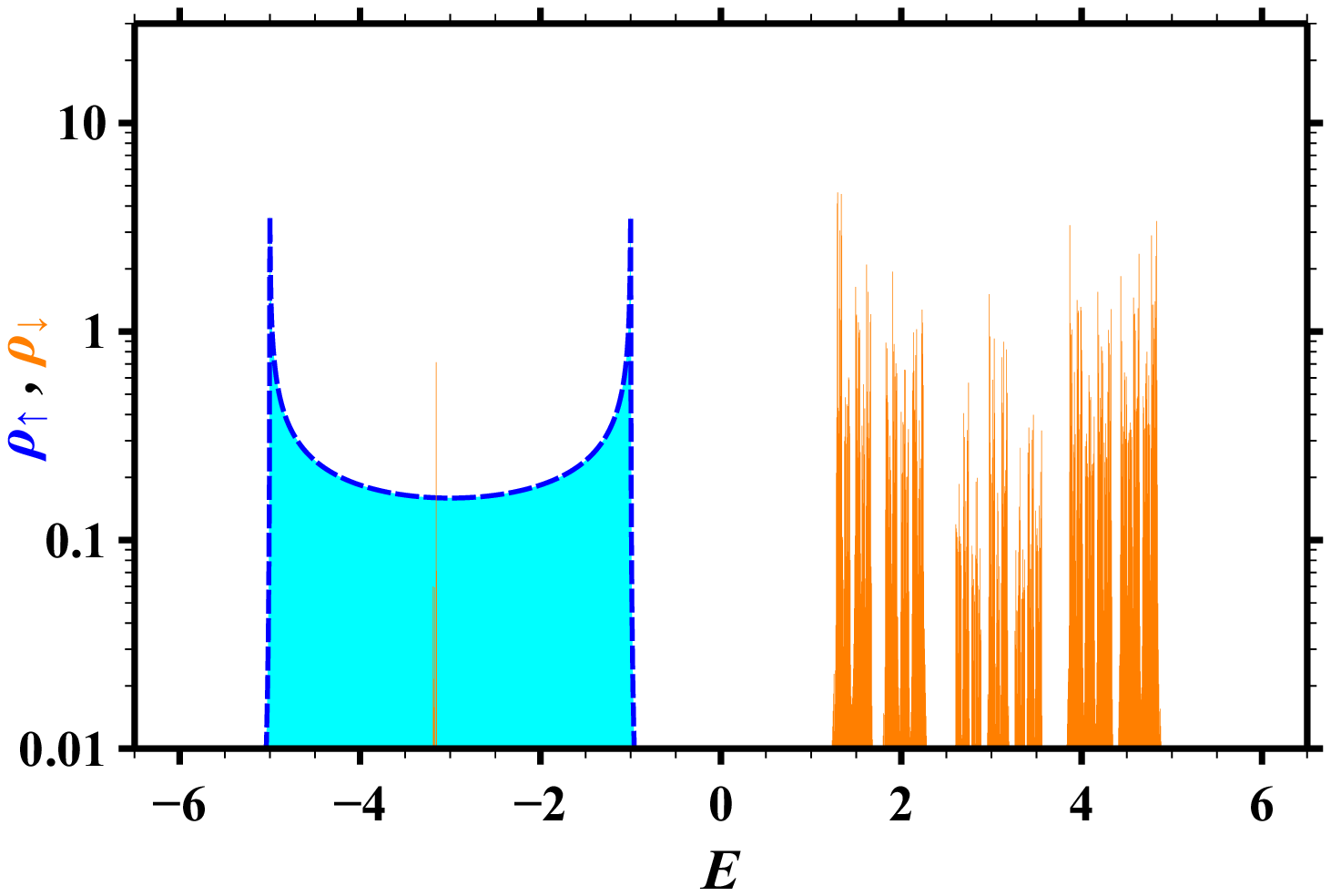}
(b)\includegraphics[width=0.95\columnwidth]{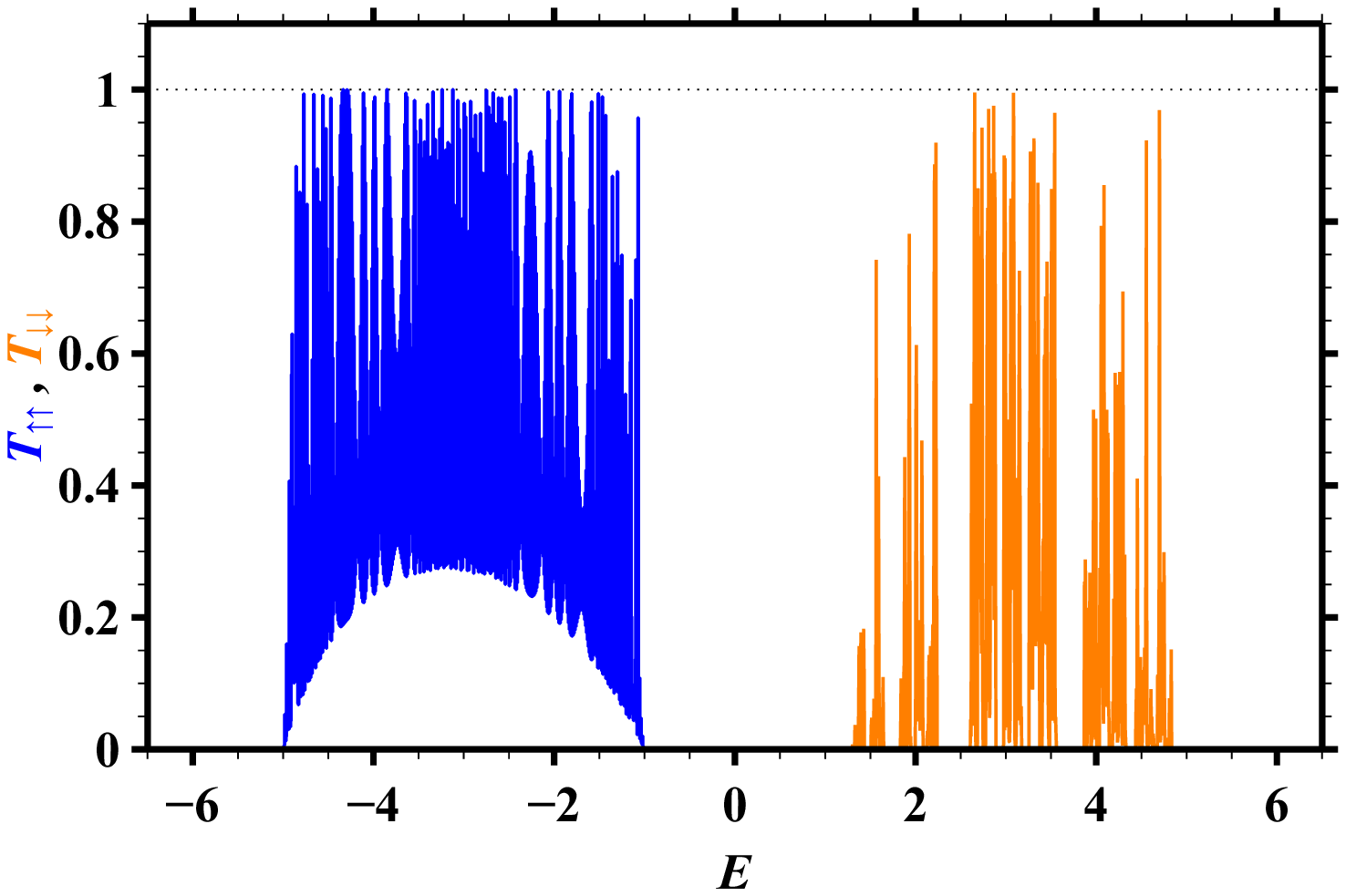}
\caption{(Color online) 
(a) DOS for spin-$1/2$ particles in the triangle-dot geometry shown in Fig.\ \ref{fig-fluxlattice} (a+b) for $\epsilon_i=0$, $h=3$, $t_L=1$, $\lambda=t_L/\sqrt{2}$ and additional magnetic flux $\Phi=\Phi_0/4$. Colors distinguishing spin-$\uparrow$ and -$\downarrow$ are as in Fig.\ \ref{fig-stub}.
(b) Transmission coefficient for the triangle geometry and parameters as in (a) with lead parameters of the non-magnetic leads given by $\epsilon_\mathrm{lead}=0$ and $t_\mathrm{lead}=3 t_L$.}
\label{fig-triangle}
\end{figure}

\section{Spin filtering triggered by an external magnetic field}

\subsection{A quasiperiodic triangle-dot array}

We now have a look at lattices shown in Fig.~\ref{fig-fluxlattice}. To gain an insight, let us focus on the simplest of them, viz, Fig.~\ref{fig-fluxlattice}(a), a quasiperiodic triangle-dot array, and its renormalized version which is the effective one dimensional Fibonacci chain shown in (b). In the array of triangles and dots a uniform magnetic flux $\Phi$ threads each triangular plaquette. The corresponding magnetic field points, say, in the positive $z$-direction. The hopping integral along an arm of the triangle is designated by $\lambda$, and it now carries a `Peierls' phase with it. The phase factor is $\pm \exp(2\pi i \Phi/3\Phi_0)$between the vertices of the triangle, $\Phi$ being the flux `trapped' in the triangle, and $\Phi_0=hc/e$ being the flux quantum. The `triangle-dot' array presents a system where the time reversal symmetry is broken, but only partially, as the particle hops along the edges of the triangle. The on-site potentials of the effective $\beta$ and $\gamma$ sites on the linear backbone become $\epsilon_{\beta,\sigma} = \epsilon_{\gamma,\sigma} = \epsilon + \lambda^2/(E-\epsilon_N)$ on decimating the non-magnetic vertices. 
\begin{figure*}[bt]
(a)\includegraphics[width=0.95\columnwidth]{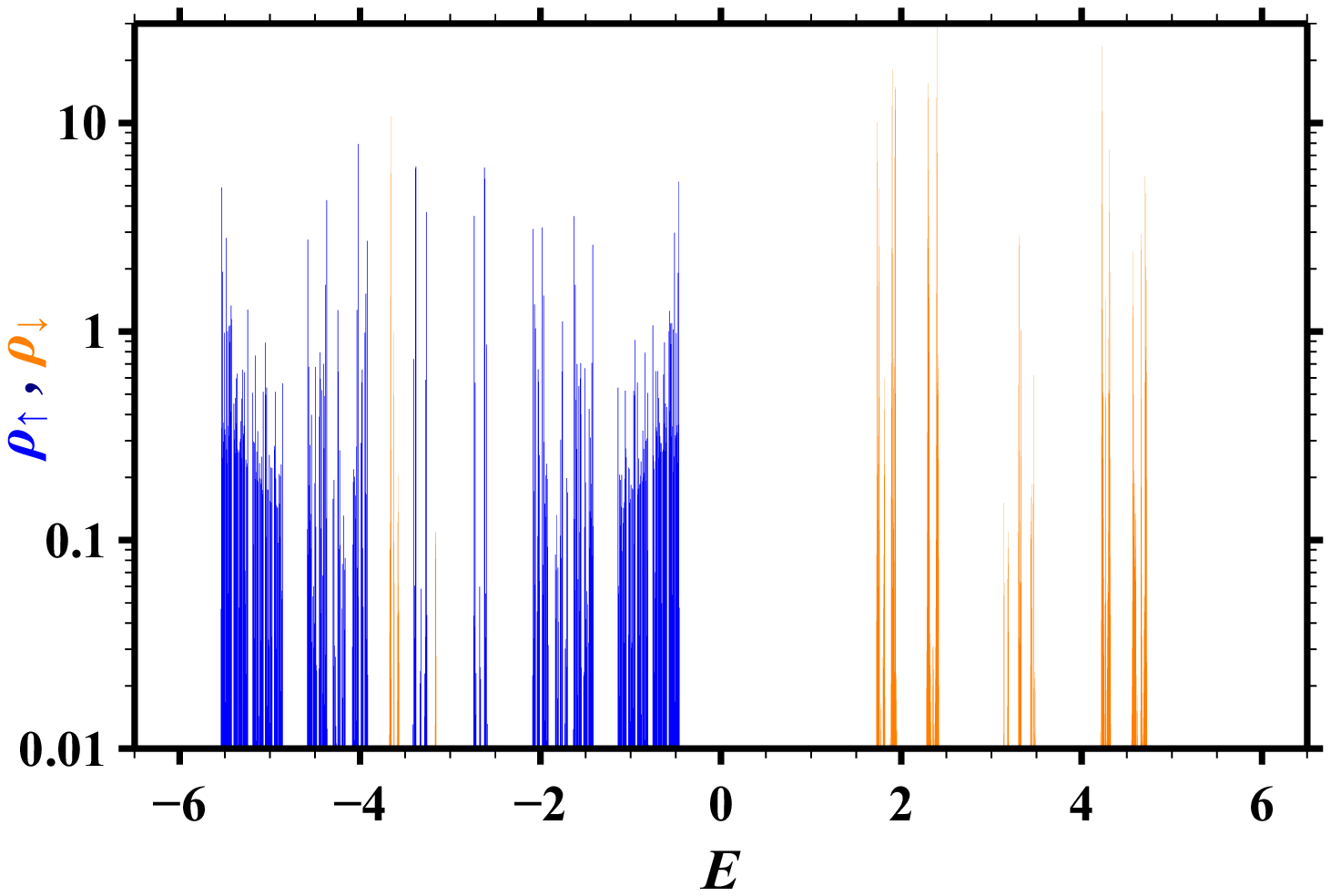}
(c)\includegraphics[width=0.95\columnwidth]{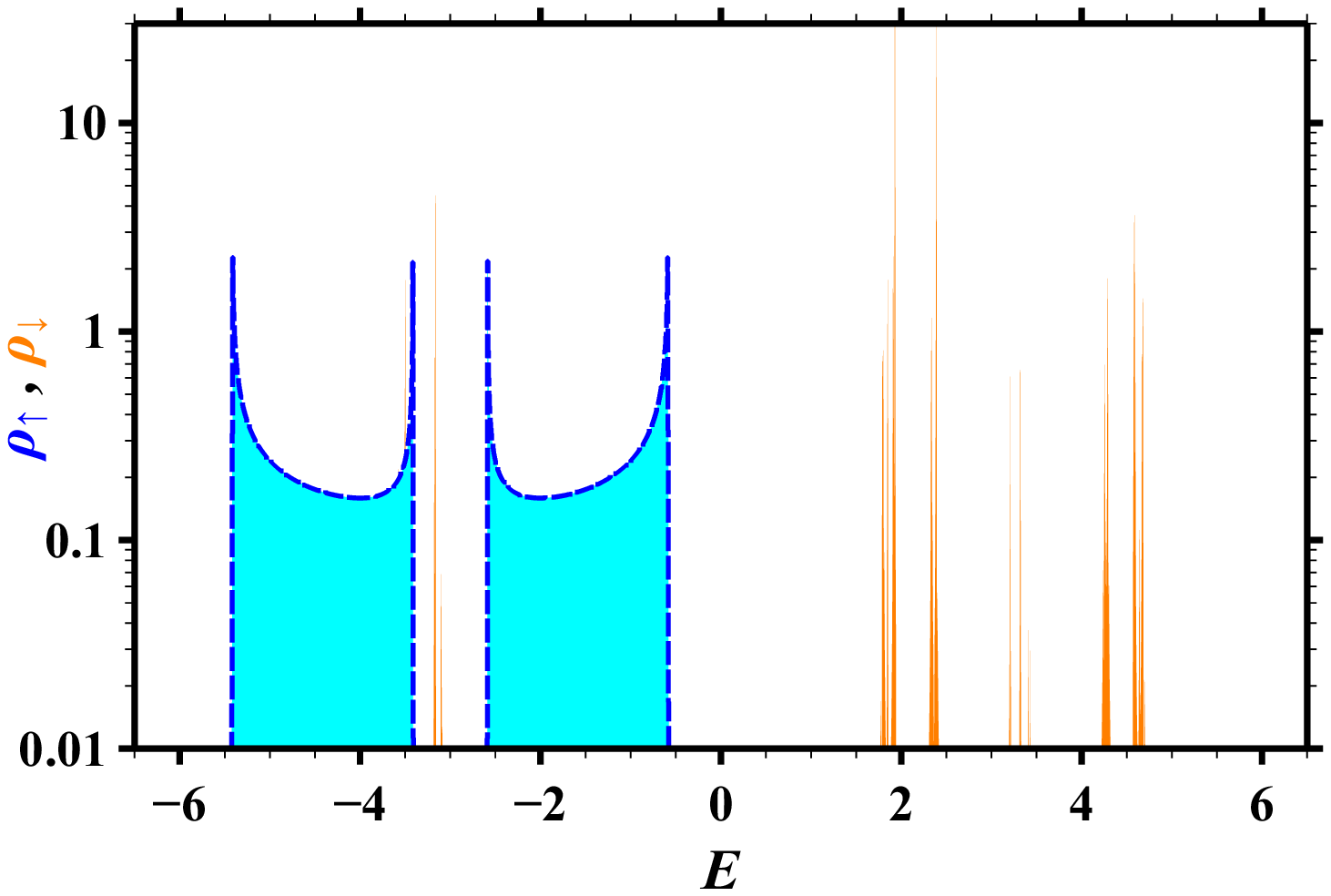}\\
(b)\includegraphics[width=0.95\columnwidth]{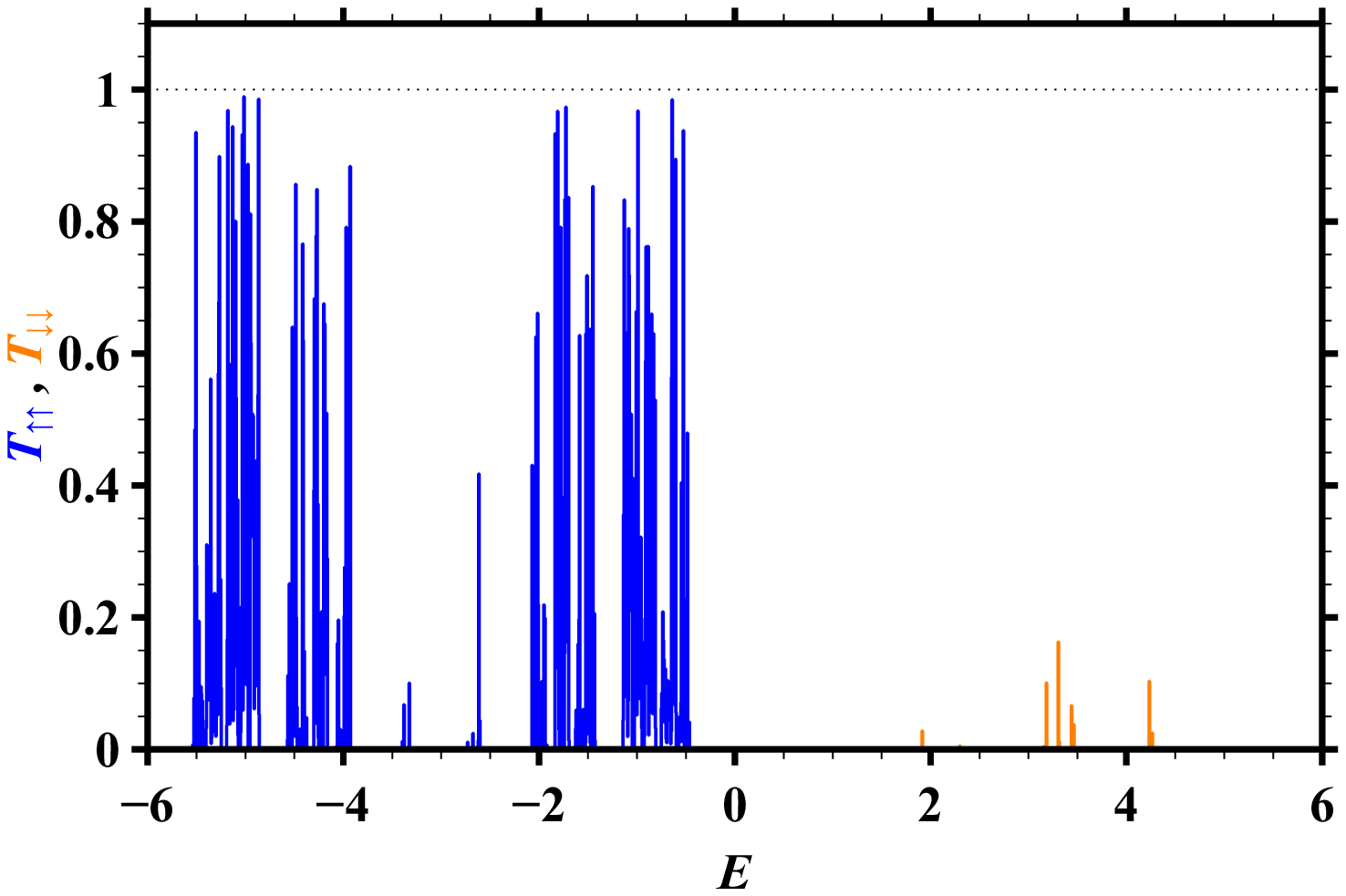}
(d)\includegraphics[width=0.95\columnwidth]{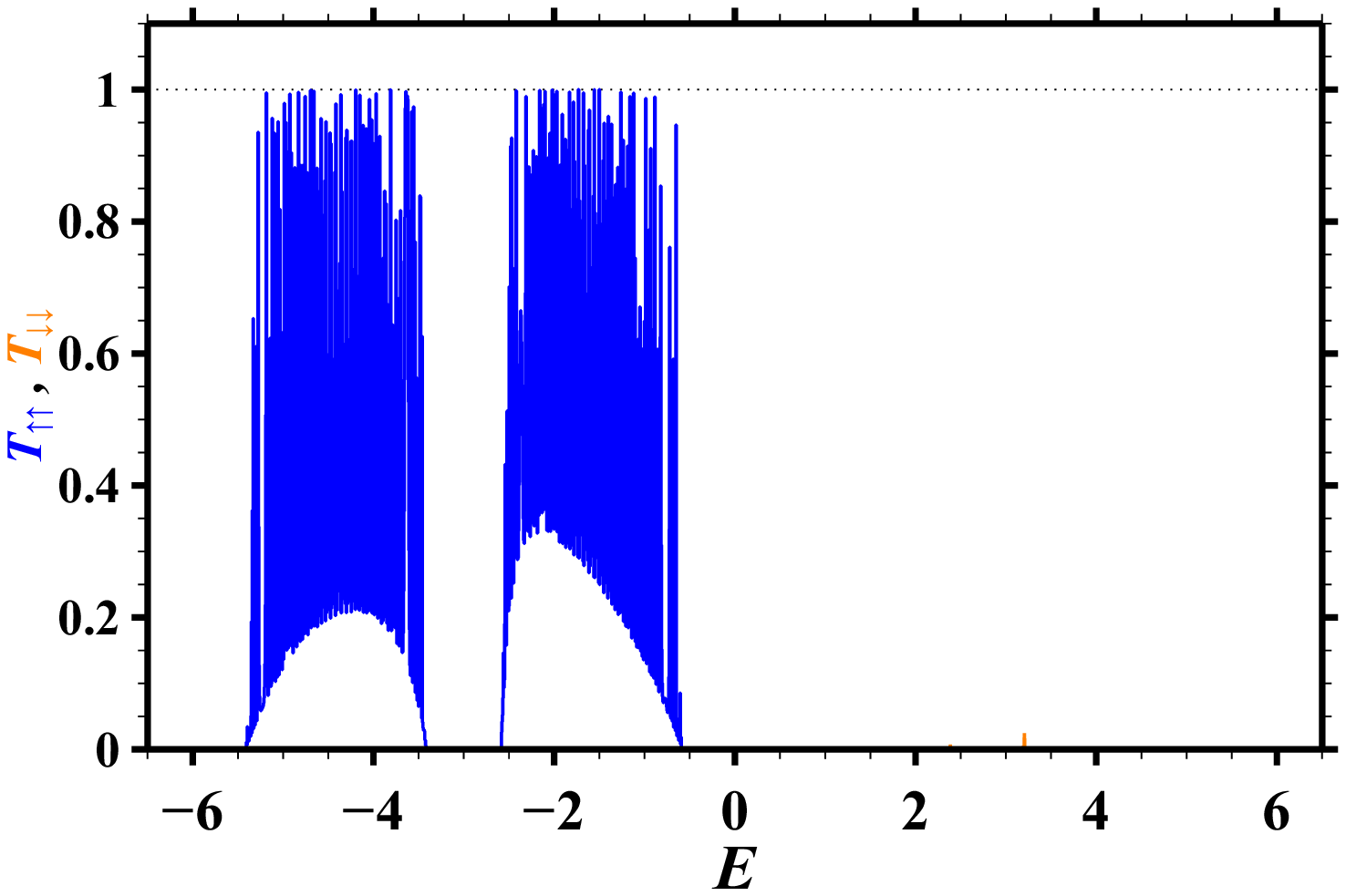}
\caption{(Color online) 
(a) DOS for spin-$1/2$ particles in the diamond-stub geometry shown in Fig.\ \ref{fig-fluxlattice} (c+d) for $\epsilon_i=0$, $h=3$, $t_L=1$, $\Phi=0$. Colors distinguishing spin-$\uparrow$ and -$\downarrow$ are as in Fig.\ \ref{fig-stub}.
(b) Transmission coefficient for the diamond-stub geometry and parameters as in (a) with lead parameters of the non-magnetic leads given by $\epsilon_\mathrm{lead}=0$ and $t_\mathrm{lead}=3 t_L$.
(c) and (d) show similar results as in (a+b), respectively, but now for external flux at $\phi=\phi_0/4$.}
\label{fig-diamond}
\end{figure*}
The hopping integral between the $\beta\gamma$ pair, as a result of the decimation of the non-magnetic vertices now depends on the energy $E$ and acquires an overall phase. The `forward' (F) and `backward' (B) hopping integrals across the $\beta\gamma$ pair, written as $t_S^{F(B)} \equiv t_{\beta (\gamma) \rightarrow \gamma (\beta)}$, are now given by, 
$t_S^{F(B)} = t_S e^{\pm i\eta}$, where
\begin{eqnarray}
t_S & = & \sqrt{\lambda^2 + \frac{\lambda^4}{(E-\epsilon_N)^2}+\frac{2 \lambda^3}{E-\epsilon_N} \cos \left( 2\pi \frac{\Phi}{\Phi_0} \right) }, \\
\tan \eta & = & \frac{(E-\epsilon_N) \sin\Theta - \lambda \sin 2\Theta}{(E-\epsilon_N) \cos\Theta + \lambda \cos 2\Theta}.
\label{phase}
\end{eqnarray}
Here, $\Theta=2\pi\Phi/(3\Phi_0)$.

Let us explain the spirit of spin filtering in this case in terms of a spin-half projectile, just as we did before. The remaining spins can be analyzed following the scheme discussed in the last section.
The Fibonacci array is, as before, composed of two bonds, characterized by the hopping integrals $t_L$ and $t_S \exp{\pm i\eta}$, along which the time reversal symmetry is broken. The decoupled set of equations for $\theta_n=\phi_n=0$, and $\sigma=\pm 1/2$ respectively, are now

\begin{eqnarray}
\left [E-(\epsilon -2\sigma h)  \right ]\psi_{n,\sigma} & = & 
t_L \psi_{n-1,\sigma} + t_L \psi_{n+1,\sigma} \nonumber \\
\left [E - \left(\epsilon -2\sigma h + \frac{\lambda^2}{E-\epsilon_N}\right) \right ] \psi_{n,\sigma} & 
= & t_L \psi_{n-1,\sigma} + 
t_S e^{i\eta} \psi_{n+1,\sigma} \nonumber \\
\left [E - \left(\epsilon -2\sigma h + \frac{\lambda^2}{E-\epsilon_N}\right) \right ] \psi_{n,\sigma} & = & 
t_S e^{-i\eta} \psi_{n-1,\sigma} + t_L \psi_{n+1,\sigma} \nonumber \\
\label{uptriangle}
\end{eqnarray}
for the $\alpha$, $\beta$ and $\gamma$ sites respectively.

%
Following the reasoning given before, let us choose the first subset of these equations, and set $\epsilon_N=\epsilon-h$ (for $\uparrow$ spins). The commutator $[\mathcal{M}_{\alpha,\uparrow},\mathcal{M}_{\gamma\beta,\uparrow}]$ reads,
\begin{widetext}
\begin{equation}
\left[\mathcal{M}_{\alpha,\uparrow},\mathcal{M}_{\gamma\beta,\uparrow} \right] = 
\frac{e^{4 \pi i \frac{ \Phi}{3 \Phi_0}}\left [(t_L^2-2\lambda^2)(E-\epsilon+h)-2\lambda^3 \cos \left(2 \pi \frac{\Phi}{\Phi_0}\right)\right ]}{\lambda t_L [ e^{2\pi i \frac{\Phi}{\Phi_0}}
(E-\epsilon+h) + \lambda]}
\left( \begin{array}{cccc}
0 & 1 \\ 
1 & 0 
\end{array}
\right) .
\label{trianglecom}
\end{equation}
\end{widetext}
The commutator is seen to vanish irrespective of energy for $\lambda=t_L/\sqrt{2}$, and for a {\it  magnetic flux $\Phi=\Phi_0/4$}. The spectrum consists of absolutely continuous subbands, for the $\uparrow$ spins only. The transport for the $\uparrow$ spins remains perfect and unattenuated in these parts of the spectrum. The DOS for the $\downarrow$ spins exhibit the fragmented structure.  The $\downarrow$ spins exhibit very weak transport in the energy regime where the $\downarrow$ spin band presents the scanty, fragmented, typical Fibonacci like spectrum. 

The reasoning holds perfectly well for any general spin projection $\sigma$, as before. The combination of $\lambda=t_L/\sqrt{2}$ and $\Phi=\Phi_0/4$ allows just one spin channel out of the available $(2s+1)$ channels, blocking the others. Of course, with a general spin projection $\sigma$, one needs to gate the potential $\epsilon_N$ appropriately, depending on which spin channel one wants to filter out. 
Furthermore, we note that the condition $\Phi=\Phi_0/4$ follows from having the effective hoping, i.e.\ across the decimated loops enclosing the flux $\Phi$, be independent of a phase difference between clockwise and counterclockwise propagation.

In Fig.~\ref{fig-triangle} (a), we show the DOS for the $\uparrow$ and the $\downarrow$ spins, and the corresponding transmission coefficients in Fig.~\ref{fig-triangle} (b) when the conditions for commutation of the matrices is fulfilled. As expected, the transmission coefficient for the $\uparrow$ spins is high and continuously distributed precisely spanning the absolutely continuous band for the $\uparrow$ spins, shown in Fig.~\ref{fig-triangle} (a). A single spike of the $\downarrow$ spin DOS is located around the middle for the $\uparrow$ spin DOS. However, extended and localized states can not coexist at the same energy, and in the convolved DOS of the full system, the state becomes perfectly extended. This is confirmed by the plot of the transmission coefficients in panel (b).

\subsection {The diamond-stub array}


In Fig.~\ref{fig-fluxlattice} we present a few prototype systems among a variety of networks that exhibit spin filtering under the influence of an external magnetic field. While (a) and (c) in the figure need a different tunnel hopping amplitude $\lambda$ in comparison to $t_L$ in the backbone, thus requiring an engineering of the hopping amplitude itself, the network shown in Fig.~\ref{fig-fluxlattice} (e), a quasiperiodic Fibonacci array of diamonds and stubs can serve the purpose with a uniform hopping integral $t_L$ throughout, including the hopping along the arms of the diamonds. We choose to discuss it explicitly, presenting the commutator for Fig.~\ref{fig-fluxlattice}(c) in the appendix. 
It may be mentioned that a similar diamond quantum network, but without the stubs, in a periodic array was considered recently to study the spin polarized transport within a tight binding framework.\cite{Pal2016f}
Furthermore, we have also investigated more complex situations in which the flux-enclosing loop contains more sites than the maximally four shown in Fig.~\ref{fig-fluxlattice}. In all such situations, a similar spin-filtering effect can be found.

Let us fix $\epsilon_N=\epsilon-h$, keeping in mind, that we are interested in filtering out the $\uparrow$ spin for $s=1/2$. In addition, we set $\lambda=t_L$. On decimating the non-magnetic vertices in Fig.~\ref{fig-fluxlattice}(e), the so called `short' hopping in the resulting Fibonacci chain becomes equal to $t_S= 2t_L^2 \cos(\pi\Phi/\Phi_0)/(E-\epsilon_N)$. The spin filtering can be effected in this case by {\it tuning the external magnetic field alone} threading every diamond plaquette. This can be quite interesting from the standpoint of an experiment. We provide the commutation conditions for Fig.~\ref{fig-fluxlattice}(c) in appendix, and give the explicit results for the last case, which is the so called `diamond-stub' case, as presented in Fig.~\ref{fig-fluxlattice}(e).
The commutation of the matrices, once again, talking in terms of the $\uparrow$ spin filtering in the spin-$1/2$ case is given by, 
\begin{equation}
\left[\mathcal{M}_{\alpha,\uparrow},\mathcal{M}_{\gamma\beta,\uparrow} \right] = 
-\frac{t_L \cos \left(2 \pi \frac{ \Phi}{\Phi_0}\right) \sec\left( \pi\frac{\Phi}{\Phi_0}\right)}{E-\epsilon+h}
\left( \begin{array}{cccc}
0 & 1  \\ 
1 & 0 
\end{array}
\right)
\label{diadotcomm}
\end{equation}
It is easily seen that, the commutator vanishes for $\Phi=\Phi_0/4$. 

It is equally important to ensure again that, as the commutation condition is satisfied, the spectrum of a periodic array of diamonds --- equivalent to the array of a $\beta\gamma$ doublet in Fig.~\ref{fig-fluxlattice}(f) --- and the spectrum of a periodic array of the $\alpha$ sites merge. In this way, one gets a perfect spin filter over a unique span of energy, for the $\uparrow$ or the $\downarrow$ spins. The range of energy of course, depends on whether we set $\epsilon_N=\epsilon-h$, or $\epsilon+h$. 
%
We have checked it in this case also. Let us write , for $\epsilon_N=\epsilon-h$, the local DOS for a periodic $\alpha$-chain as $\rho_{\alpha,\uparrow}=1/(\pi \sqrt{Q_1})$, and that of a periodic chain of $\beta\gamma$ doublet as $\rho_{\beta\gamma,\uparrow} = 1/(\pi \sqrt{Q_2})$. It is easy to work out that the difference $\Delta \equiv Q_1 - Q_2$ is given by
\begin{equation}
\Delta = \frac{4 t_L^4 F(E,\Phi)}{(E-\epsilon+h)^2 [(E-\epsilon+h)^2 - 2 t_L^2]^2} \cos \left (2 \pi \frac{\Phi}{\Phi_0}\right )
\label{diadiff}
\end{equation}
where, $F(E,\Phi)=E^3 [E-4(\epsilon-h)]+(\epsilon-h)^2 [(\epsilon-h)^2 - 3t_L^2] + 2t_L^4 + 3 E^2 [2 (\epsilon-h)^2 - t_L^2] + 2E(\epsilon-h)[3 t_L^2 -2 (\epsilon-h)^2] - t_L^4 \cos(2\pi\Phi/\Phi_0)$. 
It is clearly observed that, as soon as we set $\Phi=\Phi_0/4$, the difference $\Delta$ becomes equal to zero, and the DOS' merge. 
This happens for all the geometries discussed in this work, if we include the appropriate correlations in the numerical values of the potentials and the hopping elements $\lambda$ and $t$, where applicable. The summary is, if we fix the dot potential $\epsilon_N = \epsilon \mp h$ at the very outset, then a perfect spin filter for the $\uparrow$ or the $\downarrow$ spin electrons can be achieved by tuning the magnetic flux alone. Needless to say, that the scheme works equally well for any arbitrary spin $s$. The appropriate selection of $\epsilon_N$ will have to be made at the beginning of the experiment. The rest can be achieved simply by tuning the flux.

The DOS and the transmission coefficients for the $\uparrow$ and $\downarrow$ spins in the spin-$1/2$ case are shown in Fig.~\ref{fig-diamond} in four panels. For comparison, we show the `off resonance' condition, with $\Phi=0$, and the `resonance' condition with $\Phi=\Phi_0/4$ in separate pairs of panels (a), (b) and (c), (d) respectively. The transmission coefficient $T_{\uparrow,\uparrow}$ for the $\uparrow$ spin retains the fractal distribution, while the $\downarrow$ spins are practically forbidden even in a $377$ bonds long lattice. With $\Phi=\Phi_0/4$, the transmission of the $\downarrow$ spins is totally blocked, as is evident from Fig.~\ref{fig-diamond} (d). 

\section{Conclusions}
\label{sec-conclusions}

We have analyzed the prospect of filtering out any arbitrary spin state and letting it tunnel through an infinitely long array of quasi-one dimensional \emph{quantum networks}, arranged in an aperiodic fashion. The scheme, valid for any random or deterministically disordered arrangement of the network units, relies on opening up of subtle, hidden dimensions, $(2s+1)$ in number, to an incoming particle of spin $s$. The correlations between the values of the potential and the tunnel hopping integrals needed to filter out a specific spin channel \emph{for all energies}, and blocking the other channels, are discussed in detail. In another set of lattice structures, it has been discussed how a spin filtering effect can be observed by using an external magnetic field alone. This last issue may present an interesting experimental challenge in terms of novel spin controlled devices. The method outlined here is likely to be applicable to some  photonic structures, developed recently using ultrafast laser inscription.\cite{Mukherjee2015} One can thus look forward to engineer a `polarization filter' for photons even. Work in this direction is in progress.

\acknowledgments
The work has been supported by UGC, India and the British Council through UKIERI, Phase III, bearing reference numbers F.\ 184-14/2017(IC) and UKIERI 2016-17-004 in India and the U.K., respectively. A.M.\ is thankful to DST, India for an INSPIRE fellowship [IF160437] provided by the DST, India. Both A.M.\ and A.C.\ gratefully acknowledge the hospitality of the University of Warwick where this work was completed. Illuminating conversation with Sebabrata Mukherjee is thankfully acknowledged. UK research data statement: All data accompanying this publication are directly available within the publication.

\appendix

\section{Diamonds and dots at zero flux}

We refer to Fig.~\ref{fig-lattice1}(c) and (d). On the renormalized lattice (d), the on site potentials for the spin-half case, and the hopping integrals are given by,
\begin{eqnarray}
\epsilon_{\alpha,\sigma} & = & \epsilon - 2\sigma h \nonumber \\
\epsilon_{\beta,\sigma} & = & \epsilon_{\gamma,\sigma} = \epsilon - 2\sigma h + \frac{2\lambda^2}{E-\epsilon_N}
\label{diadot}
\end{eqnarray}
The nearest neighbor hopping integral is $t_L$ for the long bond as before, while across the `short' bond we now have $t_S=2 \lambda^2/(E-\epsilon_N)$. 
The difference equations read,
\begin{eqnarray}
\left[E-(\epsilon - 2\sigma h )\right]\psi_{n,\sigma} & = & 
t_L \psi_{n-1,\sigma} + t_L \psi_{n+1,\sigma} \nonumber \\
\left [ E - \left (\epsilon - 2\sigma h +\frac{2\lambda^2}{E-\epsilon_N} \right) \right ] \psi_{n,\sigma} & = & t_L \psi_{n-1,\sigma} + 
t_S \psi_{n+1,\sigma} \nonumber \\
\left[ E - \left(\epsilon - 2\sigma h + \frac{2\lambda^2}{E-\epsilon_N} \right) \right ] \psi_{n,\sigma} & = & t_S \psi_{n-1,\sigma} + t_L \psi_{n+1,\sigma} \nonumber \\
\label{diadoteqnup}
\end{eqnarray}
In each set the sequence of equations, from top to bottom, represents the $\alpha$, $\beta$ and the $\gamma$ sites respectively.

The transfer matrices $\mathcal{M}_{\alpha,\uparrow}$ and 
$\mathcal{M}_{\gamma\beta,\uparrow}$ $\equiv$ 
$\mathcal{M}_{\gamma,\uparrow}$ $\mathcal{M}_{\beta,\uparrow}$ can now easily be constructed following the old prescription, and the commutator, for the $\uparrow$ spin, for example, becomes,  
\begin{equation}
\left[\mathcal{M}_{\alpha,\uparrow},\mathcal{M}_{\gamma\beta,\uparrow} \right] = 
\frac{E (t_L^2-2\lambda^2)+2\lambda^2(\epsilon-h)-\epsilon_Nt_L^2}{2t_L\lambda^2}
\left( \begin{array}{cccc}
0 & 1 \\ 
1 & 0 
\end{array}
\right)
\label{com}
\end{equation}
The off diagonal elements, and hence the entire commutator vanishes for $\lambda=t_L/\sqrt{2}$, and $\epsilon_N=\epsilon-h$.

\section{Hexagon and stub in zero flux}

This geometry needs all the nearest neighbor hopping integrals to be identical to see the desired spin filtering effects. We take every  nearest neighbor hopping integral along the backbone, including the arms of the hexagon and the backbone - stub atom  tunnel hopping $\lambda$ equal to $t_L$. For spin projection $\sigma$, on renormalization the on-site potentials and the hopping integrals along the effective one dimensional Fibonacci chain in Fig.~\ref{fig-lattice1}(f) read, 
\begin{eqnarray}
\epsilon_{\alpha,\sigma} & = & \epsilon-2\sigma h+\frac{t_L^2}{E-\epsilon_N} 
\nonumber \\
\epsilon_{\beta,\sigma} =\epsilon_{\gamma,\sigma} & = & \epsilon-2\sigma h+\frac{2 (E-\epsilon_N)t_L^2}
{(E-\epsilon_N)^2 - t_L^2} \nonumber \\
t_S & = & \frac{2t_L^3}{(E-\epsilon_N)^2 - t_L^2}
\label{hexastub}
\end{eqnarray}
The commutator that we are interested in reads, 
\begin{widetext}
\begin{equation}
\left[\mathcal{M}_{\alpha,\uparrow},\mathcal{M}_{\gamma\beta,\uparrow} \right] = 
\frac{(\epsilon-h-\epsilon_N)}{2t_L^2 (E-\epsilon_N)} 
\left[(E-\epsilon_N)^2+t_L^2 \right]
\left( \begin{array}{cccc}
0 & 1 \\ 
1 & 0 
\end{array}
\right)
\label{hexacom}
\end{equation}
\end{widetext}
which clearly vanishes as we set $\epsilon_N=\epsilon-h$.

\section{The array of square networks threaded by a magnetic flux}

We now provide with the commutator for the geometry depicted in Fig.~\ref{fig-fluxlattice}(b). The squares can stand isolated, as well as can touch each other, as shown. The other cases, including any general spin $s$ situation can be worked out easily.

The effective on site potentials and the hopping integrals, for a spin projection $\sigma (=\pm 1/2)$ are given by, 
\begin{eqnarray}
\tilde\epsilon_{\alpha,\sigma} & = & \epsilon-2\sigma h + \frac{2 \lambda^2 (E-\epsilon_N)}{(E-\epsilon_N)^2 - \lambda^2} \nonumber \\
\tilde\epsilon_{\beta,\sigma} = \tilde\epsilon_{\gamma,\sigma} & = & \epsilon-2\sigma h + \frac{\lambda^2 (E-\epsilon_N)}{(E-\epsilon_N)^2 - \lambda^2} \nonumber \\
t_L^F & = & \lambda e^{i\Theta} + \frac{\lambda^3 e^{-3i\Theta}}{(E-\epsilon_N)^2-\lambda^2}
\label{square}
\end{eqnarray}
Here, $\Theta=\pi \Phi/2\Phi_0$. The double bonds are the `short' bonds in our description of a Fibonacci sequence, and has the hopping integral $t_S$ associated with it, while the hopping along the `long' bonds is now associated with a phase, as is obvious from Eq.~\eqref{square}.

For the $\uparrow$ spin, the construction of the matrices $\mathcal{M}_{\alpha,\uparrow}$ and 
$\mathcal{M}_{\gamma\beta,\uparrow} \equiv \mathcal{M}_{\gamma,\uparrow} \mathcal{M}_{\beta,\uparrow}$ are now straightforward. To simplify matters, let us preset $\epsilon_N=\epsilon-h$. The commutator, for a given set of values of $t_S$, $\epsilon$ and $h$, and $\sigma=1/2$ now reads
\begin{equation}
\left[\mathcal{M}_{\alpha,\uparrow},\mathcal{M}_{\gamma\beta,\uparrow} \right] = 
\xi
\left( \begin{array}{cccc}
0 & m_{12} \\ 
m_{21} & 0 
\end{array}
\right)
\label{square}
\end{equation}
where, 
\begin{equation}
\xi = 2\lambda^4 \cos \left(2 \pi\frac{\Phi}{\Phi_0} \right ) + (t_S^2-2\lambda^2) \left[\lambda^2 - (E-\epsilon+h)^2  \right] \\
\end{equation}
and 
\begin{widetext}
\begin{eqnarray}
m_{12} & = & e^{3\pi i\frac{\Phi}{2\Phi_0}} \frac{\left[(E-\epsilon+h)^2 e^{-i\pi \frac{\Phi}{\Phi_0}} + 2 i\lambda^2 \sin( \pi\frac{\Phi}{\Phi_0})\right ]}{\lambda t_S \left [2i\lambda^2 \sin (\pi\frac{\Phi}{\Phi_0}) + e^{\pi i\frac{\Phi}{\Phi_0}} (E-\epsilon+h)^2 \right]^2} \nonumber \\
m_{21} & = & \frac{-e^{-\pi i\frac{\Phi}{2\Phi_0}}} {\lambda t_S \left [2i \lambda^2 \sin(\pi\frac{\Phi}{\Phi_0}) - e^{\pi i\frac{\Phi}{\Phi_0}} (E-\epsilon+h)^2 \right]}
\end{eqnarray}
\end{widetext}
It is easy to see that $\xi=0$ for $\lambda=t_S/\sqrt{2}$, and $\Phi=\Phi_0/4$; the commutator vanishes identically.



\bibliographystyle{prsty}
\bibliography{Mendeley_Spin_transport_AM.bib}

\end{document}